\documentclass{vak2024}
\usepackage[utf8]{inputenc}
\usepackage{url}
\usepackage{hyperref}

\begin{document}
\title{Large-format photodetecting system pCam6060 with a GSENSE6060BSI CMOS detector, developed at SAO RAS and optimized for~photometric methods}
\titlerunning{Large-format photodetecting system pCam6060}  
\author{I.~Afanasieva\inst{1}\and V.~Murzin\inst{1}\and V.~Ardilanov\inst{1}\and N. Ivaschenko\inst{1}\and M. Pritychenko\inst{1}}
\authorrunning{Afanasieva et al.} 
	%
	%
\institute{Special Astrophysical Observatory of the Russian Academy of Sciences, Nizhny Arkhyz, 369167, Russia}

\abstract{
The pCam6060 photodetecting system was developed at SAO RAS and is based on the GSENSE6060BSI photodetector manufactured by GPixel (China) with a frame format of $6144\times6144$ active pixels and a pixel size of $10\,\mu$m. The readout speed reached 11 fps. The back-illuminated detector has a wide spectral range of 200--1040 nm with a minimum quantum efficiency (QE) of 10\% and maximum sensitivity of~95\% at 580 nm. The quantum efficiency in the near-infrared range was 58\% at 850 nm. The pCam6060 system controller implements a mode of simultaneous image readout via two 12-bit video channels with different gain and their subsequent combination in the controller into a single frame with an extended 16-bit dynamic range. This method simultaneously achieves a low readout noise level (about 3~e$^{-}$) in the high-gain channel and a large dynamic range (full well capacity of about 100\,000~e$^{-}$) in the low-gain channel. Back-illuminated CMOS detectors, unlike front-illuminated devices, do not have the effect of long-term preserving the residual charge from previous exposures, which makes them suitable for recording faint objects in photometric long-exposure observation methods. Communication between the host computer and the camera was carried out via a fiber optic line at distances of up to 50 m. Video data are recorded on the computer hard drive in real-time. The pCam6060 photodetecting system is designed for astronomical applications and has a moisture-proof design.


\doi{10.26119/VAK2024-ZZZZ}
}

\maketitle
\section{Introduction}

CMOS photodetector technologies and cameras based on these technologies have long been proven in the field of astronomy~\citep{1, 2}. Modern CMOS detectors have equal CCD values for key parameters (noise, dark current), but their frame rate is ten times higher. Photometric studies in astronomy place increased demands on the characteristics of photodetecting systems, such as image registration accuracy and sensitivity, low readout noise, and high coordinate and temporal accuracy. The use of large-format high-speed CMOS photodetectors with backside illumination (BSI) significantly increases camera sensitivity, both by reducing light losses in the detector itself and by expanding the spectral range in the red region of the spectrum, which enables meeting most of these requirements~\citep{3, 4}. 

In this paper, we present an implementation of the pCam6060: photodetecting system based on a large-format GSENSE6060BSI scientific class back-illuminated (BSI) photodetector manufactured by GPixel (China) in the $6144\times6144$ ten-micron pixel format. The development, manufacturing, and laboratory tests of the camera prototype were carried out at the Special Astrophysical Observatory of the Russian Academy of Sciences (SAO RAS). We present the results of laboratory studies on the photometric characteristics of cameras and describe the image correction methods used to obtain optimal characteristics using photometric observations.

\section{Controller architecture and camera design}

The principles of building CMOS controllers, methods for controlling high-speed CMOS photodetectors, and software design for photodetecting systems are described in detail in~\citet{5, 6}. The controller architecture, interaction of its nodes, structure, and characteristics of the system are described in~\citet{7, 8}, where the GSENSE4040 photodetector is used.

In distinction to the described implementation, the GSENSE6060BSI photodetector video data are read via 50 digital lines. The pCam6060 controller implements simultaneous image readout via two 12-bit video channels with high (HG) and low (LG) gain outputs. The HDR (High Dynamic Range) operating readout mode allows you to combine two high and low gain frames into one 16-bit frame with a linear light-signal transfer characteristic in real time. In this mode, a low level of readout noise in the high-gain channel and a large dynamic range in the low-gain channel are simultaneously achieved, which allows us to display details of both bright and weak objects. The set of supply voltages was also changed to suit the new photodetector~\citep{5}. Communication between the host computer and the photodetecting system is organized via a fiber optic line at distances of up to 50~m. The controller has an eight-channel Ethernet interface with a performance of 40~Gbit/s, providing a full-frame readout rate of 11~fps. 

\begin{figure*}
\begin{center}
\begin{tabular}{r@{\quad}l@{\quad}l@{\quad}l@{\quad}l@{\quad}l}
\includegraphics[width=0.4\textwidth]{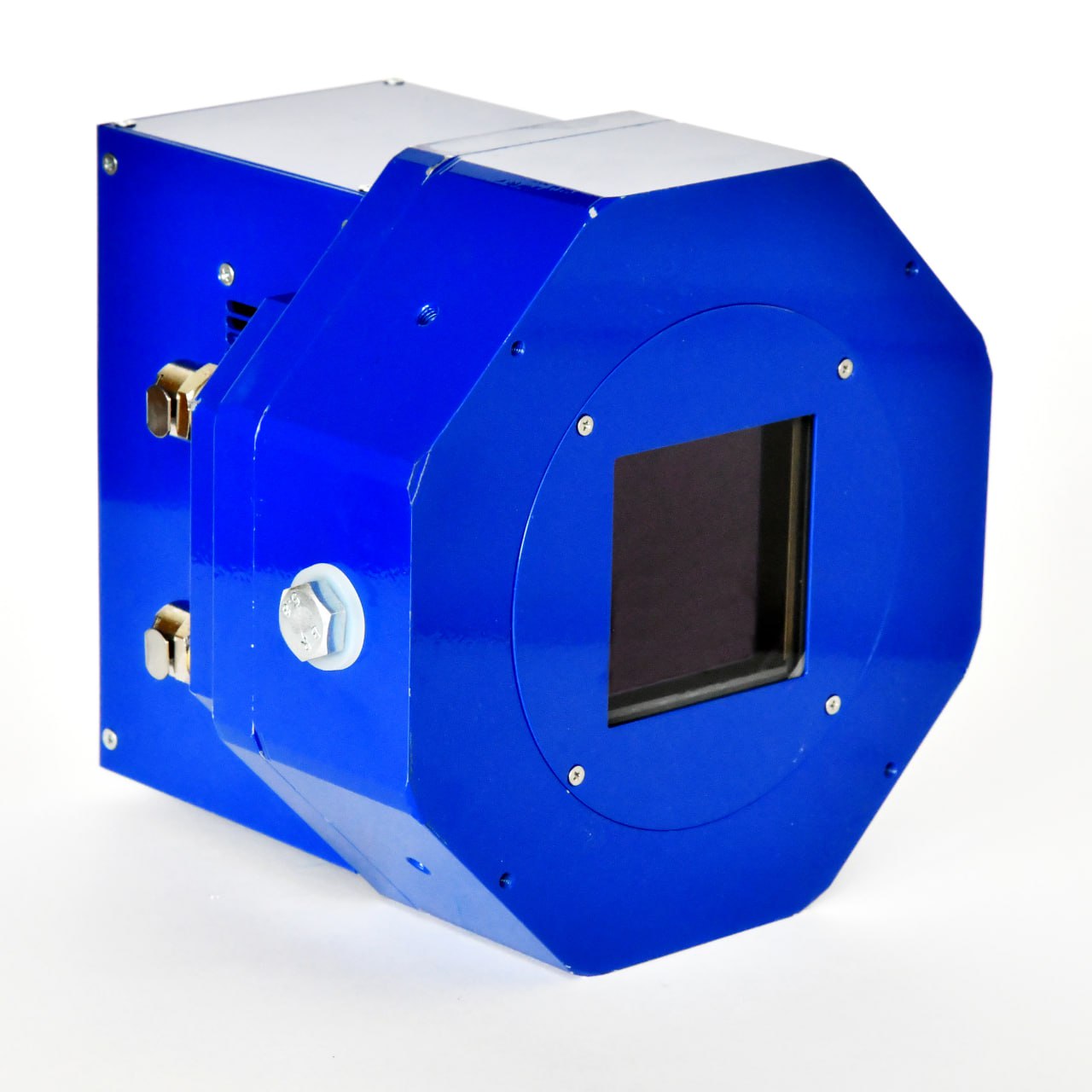}
& & & & &
\includegraphics[width=0.35\textwidth]{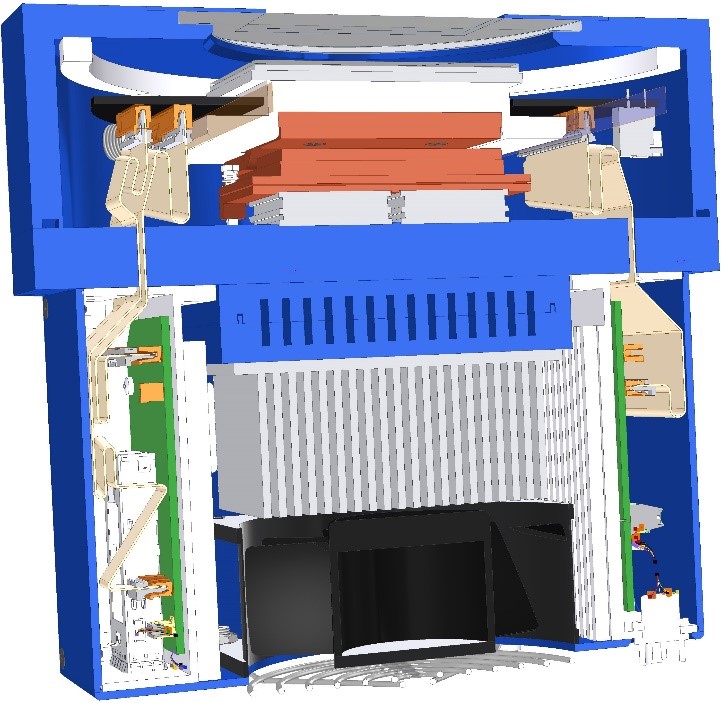}
\end{tabular}
\end{center}
\caption{Exterior view of the pCam6060 photodetecting system (left) and and the design scheme of the internal structure of the vacuum chamber of pCam6060 (right).}
\label{fig_01}
\end{figure*}

The GSENSE6060BSI-based camera is shown in Fig.~\ref{fig_01} (left). The internal structure of the optical camera is illustrated in Fig.~\ref{fig_01} (right). The camera has an optical head and housing with electronic components. To reduce heat transfer in the optical head, the photodetector is placed in an inert gas medium. The photodetector is cooled using a set of two-stage Peltier elements, which makes it possible to reach an operating temperature of $-60$\textdegree~C below the temperature of the external heat exchanger radiator. The radiator design allows air and liquid cooling. The controller’s electronics are housed in two moisture-proof housings. On the back of the chamber opposite the optical input, there are connectors for the power supply, external synchronization, and interface cable.

The camera was attached to the optical equipment via side mounting holes. If necessary, the camera can be equipped with a lens adapter. The small dimensions of the camera (190×190$\times$170 mm) make it possible to minimize light loss when placing the camera in the input light beam of the telescope.


Control and data acquisition from the GSENSE6060BSI-based camera were performed using a specially developed software package for Windows Server 2012 R2. The control program allows users to set photodetector modes, control exposure, view, analyze, process the received images, and monitor telemetry information.

\section{Results of laboratory tests}

Laboratory measurements were carried out on a flat field stand at a detector temperature of $-30$\textdegree~C. The photometric characteristics of the pCam6060 system for the three readout modes (LG, HG, HDR) are summarized in Table~\ref{tab_01}.

\begin{table}
\caption{Photometric characteristics of the pCam6060.} 
\label{tab_01}
\centering
\begin{tabular}{lrc@{\quad}c@{\quad}cl@{\quad}c@{\quad}c@{\quad}@{\quad}c}
\hline
 & & \multicolumn{3}{c}{\textbf{Measured Values}} & & \multicolumn{3}{c}{\textbf{Datasheet Values}} \\
Parameter & \textbf{Mode}  & LG & HG & HDR & & LG & HG & HDR \\
\hline
Gain factor (e$^{-}$/ADU)  &    &  $23.8$  &    $2.39$  &  $1.43$  &  &  $24.2$   &   $2.42$   &   $-$ \\ 
\multicolumn{2}{l}{Readout noise, rms (e$^{-}$)} & 23.6 & 3.47 & 3.19 &  & $22...30$ & $3...4$ & $-$ \\ 
Full well capacity (e$^{-}$) &    &  $93\,200$  &    $9\,300$  &  $91\,500$  &  &  $83\,000...95\,000$   &   $7\,000...9\,500$   &   $-$ \\ 
Dynamic range (dB)  &    &  $71.9$  &    $68.6$  &  $89.1$  &  &  $73$   &   $70$   &   $90$ \\ 
Non-linearity (\%)  &    &  $0.63$  &    $0.80$  &  $0.69$  &  &  $0.5...1.0$   &   $0.8...1.0$   &   $-$ \\ 
\hline
\multicolumn{2}{l}{Photoresponse non-uniformity (\%)} & \multicolumn{3}{c}{0.5} &  & \multicolumn{3}{c}{$0.7...1.5$}  \\ 
Gain instability (\%) & & \multicolumn{3}{c}{$0.064$} &  & \multicolumn{3}{c}{$-$}  \\ 
Image lag (e$^{-}$/pixel) & & \multicolumn{3}{c}{$3...4$} &  & \multicolumn{3}{c}{$<2$}  \\ 
\multicolumn{2}{l}{Dark current (e$^{-}$/s/pixel)} & \multicolumn{3}{c}{$0.2...0.3$} &  & \multicolumn{3}{c}{$0.04$}  \\ 
\hline
\end{tabular}
\end{table}

The GSENSE6060BSI detector does not have the long-term residual bulk image effect characteristic of FSI detectors~\citep{9,10}. However, there was a small lag: when the detector pixels were illuminated to saturation in the previous frame, in the next frame after reset, a charge of about 3~e$^{-}$ remains in these pixels.

In photometry methods, the stability of the transfer characteristics of the video channel is important. We revealed a significant dependence of the video channel gain on the detector temperature: 0.32\% on one degree of Celsius. Therefore, to obtain the required accuracy, it is necessary to precisely stabilize the detector temperature. The controller stabilizes the photodetector temperature within $\pm0.1$\textdegree~C, and the gain instability does not exceed 0.064\%.


To reduce the dark current during long exposures, the device has the following modes: lowering the pixel supply voltage and using the low-power output amplifier mode. Both of these modes were implemented in the controller. Nevertheless, the dark current we measured at long exposures remained at a fairly high level: 0.2~e$^{–}$/s/pixel at $-30$\textdegree~C. The reasons for this are currently being investigated. It is possible that this is due to the degree of quality of the detector (Grade 2).


The CMOS system with a maximum QE of 95\% at 580~nm is more powerful than the GSENSE4040-based system with a maximum QE of 74\% at 600~nm~\citep{8}.

\section{Methods of image correction}


In the raw images obtained from the CMOS detector, geometric noise is caused by interference from the operation of the control signals of the multiplexer and heterogeneity of the readout channels. In addition, we revealed the presence of a significant number of pixels in the frame for LG mode, for which the bias value was significantly higher than the average value for the frame. 

\begin{figure*}
\begin{center}
\begin{tabular}{r@{\quad}l@{\quad}l@{\quad}l@{\quad}l@{\quad}l}
\includegraphics[width=0.4\textwidth]{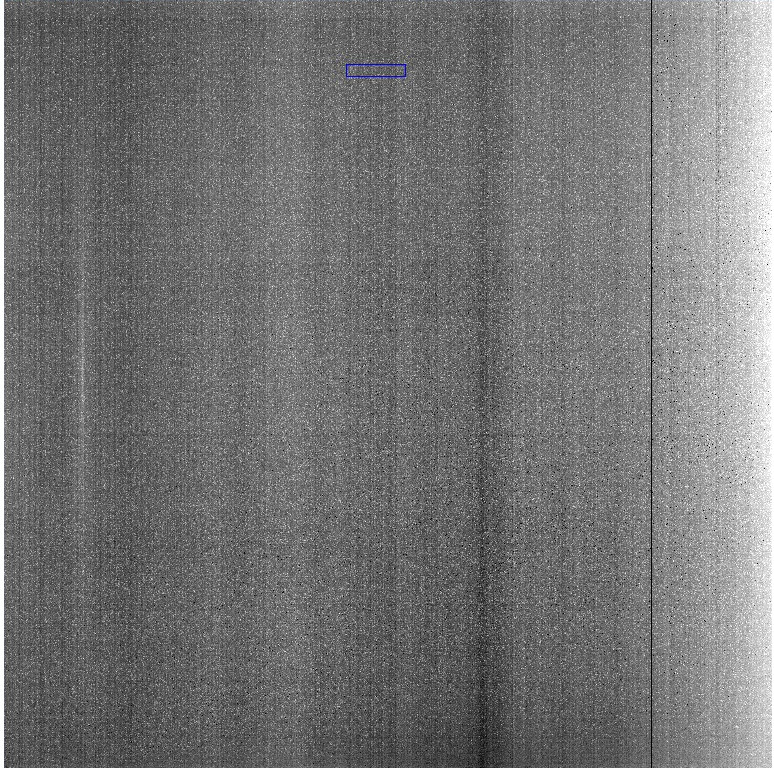}
& & & & &
\includegraphics[width=0.4\textwidth]{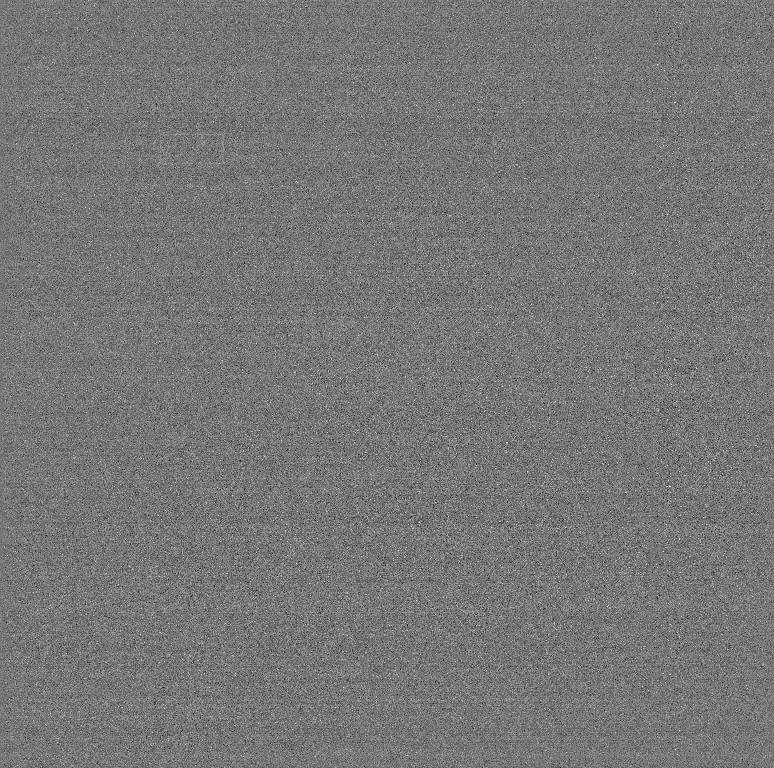}
\end{tabular}
\end{center}
\caption{The original (left) and corrected (right) bias frames in HDR mode.}
\label{fig_03}
\end{figure*}

To reduce geometric noise, the controller implements a real-time method for subtracting the average bias frame from each read frame (Fig.~\ref{fig_03}). This made it possible to reduce the readout noise in the corrected frames 1.8 times in the HG mode and 12 times in the LG mode.

\begin{figure*}
\begin{center}
\includegraphics[width=0.94\textwidth]{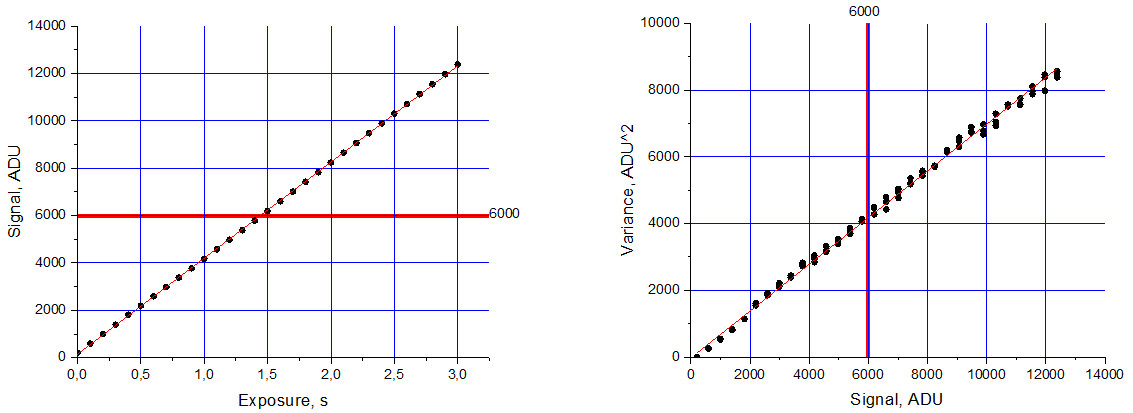}
\end{center}
\caption{Gain (left) and dispersion (right) at the junction (red line) of the LG and HG channels.}
\label{fig_04}
\end{figure*}

To couple high- and low-gain frames into an HDR frame with a linear light-signal transfer characteristic, multiplicative and additive coefficients are used, which were previously calculated based on the measured transfer characteristics of the LG and HG channels. The resulting transfer characteristics exhibit no gain (Fig.~\ref{fig_04}, left) or dispersion shift (Fig.~\ref{fig_04}, right) in the area of the junction point.

\section{Summary}

The developed large-format camera system with a frame rate of 11 fps and low readout noise is undoubtedly excellent for short-exposure observation. For use in long-exposure photometry, the proposed method implements dark-current reduction modes, bias frame subtraction, and video channel gain stabilization. In addition, to achieve the necessary accuracy, it will be necessary to use postprocessing of the obtained images --- subtraction of the dark frame, correction of photoresponse nonuniformity in each pixel over a flat field, and correction of nonlinearity.


\end{document}